\documentclass[aps,prd,superscriptaddress]{revtex4}
\usepackage{epsfig,epsf}
\usepackage{amsmath}
\usepackage{amsthm}
\usepackage{amsfonts}
\usepackage{amssymb}
\usepackage{dsfont}
\usepackage{multirow}
\usepackage{appendix}
\usepackage{slashed}
\usepackage[active]{srcltx}
\usepackage{psfrag}

\begin{document}

\title{Mass spectra of heavy hybrid quarkonia and $\overline{b}gc$ mesons}
\date{\today}
\author{A.~Alaakol}
\affiliation{Department of Physics, Marmara University, 34722 Istanbul, T\"{u}rkiye}
\author{S.~S.~Agaev}
\affiliation{Institute for Physical Problems, Baku State University, Az--1148 Baku,
Azerbaijan}
\author{K.~Azizi}
\thanks{Corresponding Author}
\affiliation{Department of Physics, University of Tehran, North Karegar Avenue, Tehran
14395-547, Iran}
\affiliation{Department of Physics, Do\v{g}u\c{s} University, Dudullu-\"{U}mraniye, 34775
Istanbul, T\"{u}rkiye}
\author{H.~Sundu}
\affiliation{Department of Physics Engineering, Istanbul Medeniyet University, 34700
Istanbul, T\"{u}rkiye}

\begin{abstract}
Masses and current couplings of the charmonium and bottomonium hybrids $
\overline{c}gc$ and $\overline{b}gb$ with spin-parities $J^{\mathrm{PC}
}=0^{++},\ 0^{+-},\ 0^{-+},\ 0^{--}$ and $1^{++},\ 1^{+-},\ 1^{-+},\ 1^{--}$
are calculated using QCD two-point sum rule method. Computations are
performed by taking into account gluon condensates up to dimension 12
including terms $\sim \langle g_{s}^{3}G^{3}\rangle ^{2}$. The parameters of
the bottom-charm hybrids $\overline{b}gc$ with quantum numbers $J^{\mathrm{PC
}}=0^{+},\ 0^{-},\ 1^{+}$, and $1^{-}$ are calculated as well. In
computations the dominance of the pole contribution to sum rule results is
ensured. It is demonstrated that all charmonia hybrids decay strongly to
two-meson final states. The bottomonium hybrids $0^{-+}$ and $1^{-+}$ as
well as the bottom-charm hybrid mesons $0^{-(+)}$ and $1^{-(+)}$ may be
stable against strong two-meson decay modes. Results of the present work are
compared with ones obtained using the sum rule and alternative approaches.
Our predictions for parameters of the heavy hybrid mesons may be useful to
study their various decay channels which are important for interpretation of
ongoing and future experiments.
\end{abstract}

\maketitle


\section{Introduction}

\label{sec:Intro}
It is well known that apart from conventional mesons and baryons, QCD and
parton model do not forbid existence of multiquark states, glueballs, and
hybrid structures. Experimental investigations of last years allowed one to
collect valuable information about tetraquark and pentaquark candidates. The
$X$ resonances presumably composed of four charm quarks (antiquarks) \cite%
{LHCb:2020bwg,Bouhova-Thacker:2022vnt,CMS:2023owd}, and structures $%
P_{c}(4330)$, $P_{c}(4440)$, and $P_{cs}(4459)$ \cite%
{LHCb:2015yax,LHCb:2019kea,LHCb:2020jpq} discovered recently are among
promising candidates to tetra- and pentaquarks, respectively.

Charmonium and bottomonium hybrids, i.e., mesons which besides valence
quarks and antiquarks contain also valence gluon(s), as well as bottom-charm
hybrids $\overline{b}gc$ belong also to a class of exotic states. There are
a few candidates to light hybrid mesons discovered in various experiments.
Namely, mesons $\pi _{1}(1400)$, $\pi _{1}(1600)$ and $\pi _{1}(2105)$ are
such states. Recently, the BESIII collaboration informed about the isoscalar
vector resonance $\eta _{1}(1855)$ with exotic quantum numbers $J^{\mathrm{PC%
}}=1^{-+}$ \cite{BESIII:2022riz}, which was observed in the process $J/\psi
\rightarrow \gamma \eta _{1}(1855)\rightarrow \gamma \eta \eta ^{\prime }$.
This resonance is considering as possible hybrid meson \cite%
{Chen:2022qpd,Qiu:2022ktc,Shastry:2022mhk} though alternative hadronic
molecule and/or diquark-antidiquark models were suggested to explain
corresponding experimental data.

There are candidates to hybrid mesons among heavy resonances as well. Thus,
it was argued that the vector resonances $\psi (4230)$ and $\psi (4360)$ may
be hybrid charmonium states $\overline{c}gc$ or their essential components
\cite{Kou:2005gt,Olsen:2017bmm}. Detailed information about numerous
resonances that may be considered as candidates to hybrid quarkonia can be
found in Ref.\ \cite{Brambilla:2022hhi}. The hybrids with baryon quantum
numbers are also among exotic hadrons. The $\Lambda (1405)$ discovered many
years ago \cite{Engler:1965zz} and investigated by various collaborations as
a candidate to such baryon \cite{Niiyama:2008rt,HADES:2012csk,CLAS:2013rxx}.

Theoretical studies of hybrid hadrons have long story: Existence of such
states was supposed more than four decades ago in Refs.\ \cite%
{Jaffe:1975fd,Horn:1977rq}. Interesting results concerning spectroscopic
parameters of hypothetical hybrid particles, their decay and production
mechanisms were obtained at early stages of investigations in the context of
different methods \cite%
{Tanimoto:1982wy,Barnes:1982tx,Chanowitz:1982qj,Isgur:1985vy,Deviron:1984svx,Govaerts:1984hc, Govaerts:1985fx,Close:1994hc,Close:1994pr,Page:1996rj,Page:1998gz,Zhu:1998sv}%
. These studies were continued in Refs.\ \cite{Narison:2009vj,Qiao:2010zh,
Harnett:2012gs,HadronSpectrum:2012gic,Chen:2013zia,Chen:2013eha,Cheung:2016bym, Azizi:2017xyx,Huang:2014hya,Palameta:2018yce,Miyamoto:2019oin, Brambilla:2018pyn,Ryan:2020iog,TarrusCastella:2021pld,Woss:2020ayi, Barsbay:2022gtu,Barsbay:2024vjt,Tang:2021zti,Chen:2022isv,Soto:2023lbh,Bruschini:2023tmm}
aimed to refine calculational schemes and methods used in relevant analyses.
The hybrid states were explored by means of QCD sum rule (SR) and lattice
methods, constituent gluon, flux-tube and nonrelativistic field theory
models which form theoretical basis of these investigations.

The wide diversity of obtained results makes relevant problems actual until
now. For instance, in Ref.\ \cite{Qiao:2010zh} the masses of the vector $%
1^{--}$ hybrids $H_{c}=\overline{c}gc$ and $H_{b}=\overline{b}gb$ were
evaluated by employing SR approach. Calculations were performed with $%
\langle g_{s}^{3}G^{3}\rangle $ accuracy and predictions $m_{Hc}=4.12-4.79~%
\mathrm{GeV}$ and $m_{H_{b}}=10.24-11.15~\mathrm{GeV}$ were made for the
masses of these particles. The masses of the same hybrids were estimated as $%
(3.36\pm 0.15)~\mathrm{GeV}$ and $(9.70\pm 0.12)~\mathrm{GeV}$ in Ref.\ \cite%
{Chen:2013zia}. The lattice simulations led to the results $(4.41\pm 0.02)~%
\mathrm{GeV}$ and $(10.95\pm 0.02)~\mathrm{GeV}$ \cite%
{Cheung:2016bym,Ryan:2020iog}. Analyses carried out in Ref.\ \cite%
{Brambilla:2018pyn} yielded $\simeq 4.40~\mathrm{GeV}$ and $\simeq 10.74~%
\mathrm{GeV}$, respectively. Recently, these problems were addressed in the
context of the Born-Oppenheimer effective field theory (BOEFT) \cite%
{Soto:2023lbh}. The authors found that $1^{--}$ states have the masses $%
(4.011\pm 0.030)~\mathrm{GeV}$ and $(10.6902\pm 0.003)~\mathrm{GeV}$,
respectively.

The exotic mesons $\overline{b}gc$ were also in the sphere of researches'
interests \cite{Chen:2013eha,Miyamoto:2019oin}. The bottom-charmonium
hybrids with quantum numbers $J^{\mathrm{P}}=0^{+}$, $0^{-}$, $1^{+}$, $%
1^{-} $, $2^{+}$, and $2^{-}$ were considered in Ref.\ \cite{Chen:2013eha}.
Calculations were carried out using QCD sum rule approach by taking into
account dimension six condensates. The masses of these hybrids were
predicted in the range of $6.8$ to $8.5~\mathrm{GeV}$. The authors analyzed
their possible strong decays including open- and hidden-flavor two-body
exclusive channels. In Ref.\ \cite{Miyamoto:2019oin} the mass spectra and
decays of the states $\overline{b}gc$ with magnetic and electric gluon were
investigated by means of the constituent gluon model.

In the present article, we compute the masses and current couplings of
hybrid quarkonia $\overline{c}gc$ and $\overline{b}gb$, and exotic mesons $%
\overline{b}gc$ with different spin-parities. Our investigations are
performed in the context of QCD two-point SR method by taking into account
nonperturbative terms proportional to $\langle g_{s}^{3}G^{3}\rangle ^{2}$.
The sum rule method is one of effective and powerful nonperturbative tools
in high energy physics. It was elaborated to study features of conventional
hadrons, and analyze their decay channels \cite%
{Shifman:1978bx,Shifman:1978by}. But this approach can be employed to
investigate exotic hadrons as well \cite%
{Nielsen:2009uh,Albuquerque:2018jkn,Agaev:2020zad}. It is remarkable that
QCD SRs was successfully applied to investigate the hybrid quarkonia
starting from first years of its invention \cite%
{Deviron:1984svx,Govaerts:1984hc,Govaerts:1985fx}.

This paper is structured in the following way. In Sec.\ \ref{sec:Hquarkonia}%
, we calculate spectral parameters of the heavy hybrids $\overline{c}gc$ and
$\overline{b}gb$ with spin-parities $J^{\mathrm{PC}}=0^{++},\ 0^{+-},\
0^{-+},\ 0^{--}$ and $1^{++},\ 1^{+-},\ 1^{-+},\ 1^{--}$. Section \ref%
{sec:Hmesons} is devoted to analysis of the hybrid mesons $\overline{b}gc$
with quantum numbers $J^{\mathrm{P}}=0^{+},\ 0^{-},\ 1^{+}$, and $1^{-}$.
The last section contains our short conclusions.


\section{Mass and current coupling of the heavy hybrid quarkonia}

\label{sec:Hquarkonia}

The sum rules for parameters of the heavy hybrid states $H^{c}=\overline{c}%
gc $ and $H^{b}=\overline{b}gb$ can be extracted from analysis of the
correlation function
\begin{equation}
\Pi _{\mu \nu }(p)=i\int d^{4}xe^{ipx}\langle 0|\mathcal{T}\{J_{\mu
}(x)J_{\nu }^{\dag }(0)\}|0\rangle .  \label{eq:CF1}
\end{equation}%
Here $J_{\mu }(x)$ is the interpolating current for the particle under
consideration, and $\mathcal{T}$ stands for a time-ordering product of two
currents.

For the scalar and vector hybrids with the negative $\mathrm{C}$-parity $J^{%
\mathrm{PC}}=0^{+-}$ and $1^{--}$ the interpolating current has the
following form%
\begin{equation}
J_{\mu }^{1}(x)=g_{s}\overline{Q}_{a}(x)\gamma ^{\alpha }\gamma _{5}\frac{{%
\lambda }_{ab}^{n}}{2}\widetilde{G}_{\mu \alpha }^{n}(x)Q_{b}(x),
\label{eq:C1}
\end{equation}%
whereas for the particles with the positive $\mathrm{C}$-parity $0^{++}$ and
$1^{-+}$ it is given by the expression%
\begin{equation}
J_{\mu }^{2}(x)=g_{s}\overline{Q}_{a}(x)\gamma ^{\alpha }\frac{{\lambda }%
_{ab}^{n}}{2}G_{\mu \alpha }^{n}(x)Q_{b}(x).  \label{eq:C2}
\end{equation}%
In Eqs.\ (\ref{eq:C1}) and (\ref{eq:C2}), the heavy quark field $Q(x)$
labels $c$ or $b$ quarks. Here $g_{s}$ is the QCD strong coupling constant, $%
a$ and $b$ are color indices and ${\lambda }^{n}$, $n=1,2,..8$ are Gell-Mann
matrices. The gluon field strength and its dual tensors are shown by $G_{\mu
\nu }^{n}(x)$ and $\widetilde{G}_{\mu \nu }^{n}(x)=\varepsilon _{\mu \nu
\alpha \beta }G^{n\alpha \beta }(x)/2$, respectively.

Interpolating currents for the pseudoscalar and axial-vector hybrids are
defined by the formulas
\begin{equation}
J_{\mu }^{3}(x)=g_{s}\overline{Q}_{a}(x)\gamma ^{\alpha }\gamma _{5}\frac{{%
\lambda }_{ab}^{n}}{2}G_{\mu \alpha }^{n}(x)Q_{b}(x),  \label{eq:C3}
\end{equation}%
in the case of the states $0^{--}$, $1^{+-}$ and
\begin{equation}
J_{\mu }^{4}(x)=g_{s}\overline{Q}_{a}(x)\gamma ^{\alpha }\frac{{\lambda }%
_{ab}^{n}}{2}\widetilde{G}_{\mu \alpha }^{n}(x)Q_{b}(x),  \label{eq:C4}
\end{equation}%
for the particles with quantum numbers $0^{-+}$ and $1^{++}$.

Let us consider the current $J_{\mu }^{1}(x)$ and charmonium hybrids $H_{%
\mathrm{S}}$ and $H_{\mathrm{V}}$ with $J^{\mathrm{PC}}=0^{+-}$ and $1^{--}$
as a sample case: Generalization to remaining currents is straightforward.
In accordance with methodology of the sum rule analysis, we first express
the correlation function $\Pi _{\mu \nu }(p)$ in terms of the particles'
physical parameters
\begin{eqnarray}
&&\Pi _{\mu \nu }^{\mathrm{Phys}}(p)=\frac{\langle 0|J_{\mu }|H_{\mathrm{S}%
}(p)\rangle \langle H_{\mathrm{S}}(p)|J_{\nu }^{\dag }|0\rangle }{m_{\mathrm{%
S}}^{2}-p^{2}}  \notag \\
&&+\frac{\langle 0|J_{\mu }|H_{\mathrm{V}}(p,\varepsilon )\rangle \langle H_{%
\mathrm{V}}(p,\varepsilon )|J_{\nu }^{\dag }|0\rangle }{m_{\mathrm{V}%
}^{2}-p^{2}}+\cdots ,  \label{eq:CF2}
\end{eqnarray}%
where $m_{\mathrm{S}}$ and $m_{\mathrm{V}}$ are the masses of the
corresponding hybrids. Here, the contributions of the ground-level hybrids $%
H_{\mathrm{S}}$ and $H_{\mathrm{V}}$ are written down explicitly, whereas
effects due to higher resonances and continuum states are denoted by the
ellipses. For simplicity of the formulas, we also replace $J_{\mu
}^{1}\rightarrow J_{\mu }$.

The Eq.\ (\ref{eq:CF2}) is derived by inserting full set of physical states
with quantum numbers of the hybrids into Eq.\ (\ref{eq:CF1}) and carrying
out integration over $x$. The expression $\Pi _{\mu \nu }^{\mathrm{Phys}}(p)$
can be further simplified by expressing the matrix elements in terms of the
masses and current couplings of $H_{\mathrm{S}}$ and $H_{\mathrm{V}}$
\begin{equation}
\langle 0|J_{\mu }|H_{\mathrm{S}}(p)\rangle=f_{\mathrm{S}}p_{\mu },\ \ \
\langle 0|J_{\mu }|H_{\mathrm{V}}(p,\varepsilon )\rangle=m_{\mathrm{V}}f_{%
\mathrm{V}}\varepsilon _{\mu },  \label{eq:ME1}
\end{equation}%
with $f_{\mathrm{S}}$ and $f_{\mathrm{V}}$ being the current couplings of
the hybrids, and $\varepsilon _{\mu }$ -- the polarization vector of $H_{%
\mathrm{V}}$.

Having substituted these matrix elements into Eq.\ (\ref{eq:CF2}), it is not
difficult to find
\begin{equation}
\Pi _{\mu \nu }^{\mathrm{Phys}}(p)=\frac{f_{\mathrm{S}}^{2}}{m_{\mathrm{S}%
}^{2}-p^{2}}p_{\mu }p_{\nu }+\frac{m_{\mathrm{V}}^{2}f_{\mathrm{V}}^{2}}{m_{%
\mathrm{V}}^{2}-p^{2}}\left( -g_{\mu \nu }+\frac{p_{\mu }p_{\nu }}{p^{2}}%
\right) +\cdots .  \label{eq:Pside}
\end{equation}%
As is seen, the correlator $\Pi _{\mu \nu }^{\mathrm{Phys}}(p)$ contains two
Lorentz structures $g_{\mu \nu }$ and $p_{\mu }p_{\nu }$, which may be
employed to derive the SRs of interest. The term proportional to $g_{\mu \nu
}$ receives contribution only from the vector particle. Therefore, the
relevant amplitude $\Pi _{\mathrm{V}}^{\mathrm{Phys}}(p^{2})$ can be safely
used to get sum rules for $m_{\mathrm{V}}$ and $f_{\mathrm{V}}$. To isolate
the contribution of the scalar state, it is convenient to multiply $\Pi
_{\mu \nu }^{\mathrm{Phys}}(p)$ by $p^{\mu }p^{\nu }/p^{2}$ which leads to
\begin{equation}
\frac{p^{\mu }p^{\nu }}{p^{2}}\Pi _{\mu \nu }^{\mathrm{Phys}}(p)=\widetilde{%
\Pi }^{\mathrm{Phys}}(p)=-f_{\mathrm{S}}^{2}+\frac{m_{\mathrm{S}}^{2}f_{%
\mathrm{S}}^{2}}{m_{\mathrm{S}}^{2}-p^{2}}+\cdots .
\end{equation}%
The correlation function after this operation contains only trivial Lorentz
structure proportional to $I$. We denote corresponding invariant amplitude
by $\Pi _{\mathrm{S}}^{\mathrm{Phys}}(p^{2})$, and use it to obtain the SRs
for parameters $m_{\mathrm{S}}$ and $f_{\mathrm{S}}$.

The QCD side of the sum rule is determined by Eq.\ (\ref{eq:CF1}) calculated
by employing the explicit expressions of the interpolating currents and
replacing contractions of heavy quark fields with relevant propagators. The
correlator obtained by this manner should be computed in the operator
product expansion ($\mathrm{OPE}$) with some accuracy. After these
operations, the correlation function $\Pi _{\mu \nu }(p)$ acquires the form
\begin{eqnarray}
&&\Pi _{\mu \nu }^{\mathrm{OPE}}(p)=\frac{i\varepsilon _{\mu \theta \alpha
\beta }\varepsilon _{\nu \delta \alpha ^{\prime }\beta ^{\prime }}}{4}\int
d^{4}xe^{ipx}\frac{{\lambda }_{ab}^{n}{\lambda }_{a^{\prime }b^{\prime }}^{m}%
}{4}\mathrm{Tr}\left[ S_{Q}^{a^{\prime }a}(-x)\gamma ^{\theta }\gamma
_{5}S_{Q}^{bb^{\prime }}(x)\gamma ^{\delta }\gamma _{5}\right]  \notag \\
&&\times \langle 0|g_{s}^{2}G^{n\alpha \beta }(x)G^{m\alpha ^{\prime }\beta
^{\prime }}(0)|0\rangle ,  \label{eq:QCD1}
\end{eqnarray}%
where $S_{Q}^{ab}(x)$ is the propagator of the $Q=c$ ($b$) quark. In present
article, we use the following expression for the propagators $S_{Q}^{ab}(x)$
\begin{eqnarray}
&&S_{Q}^{ab}(x)=i\int \frac{d^{4}k}{(2\pi )^{4}}e^{-ikx}\Bigg \{\frac{\delta
_{ab}\left( {\slashed k}+m_{Q}\right) }{k^{2}-m_{Q}^{2}}-\frac{%
g_{s}G_{ab}^{\alpha \beta }}{4}\frac{\sigma _{\alpha \beta }\left( {\slashed %
k}+m_{Q}\right) +\left( {\slashed k}+m_{Q}\right) \sigma _{\alpha \beta }}{%
(k^{2}-m_{Q}^{2})^{2}}  \notag \\
&&+\frac{g_{s}^{2}G^{2}}{12}\delta _{ab}m_{Q}\frac{k^{2}+m_{Q}{\slashed k}}{%
(k^{2}-m_{Q}^{2})^{4}}+\frac{g_{s}^{3}G^{3}}{48}\delta _{ab}\frac{\left( {%
\slashed k}+m_{Q}\right) }{(k^{2}-m_{Q}^{2})^{6}}  \notag \\
&&\times \left[ {\slashed k}\left( k^{2}-3m_{Q}^{2}\right) +2m_{Q}\left(
2k^{2}-m_{Q}^{2}\right) \right] \left( {\slashed k}+m_{Q}\right) +\cdots %
\Bigg \}.  \notag \\
&&  \label{eq:Prop}
\end{eqnarray}%
Here, we have introduced the shorthand notations
\begin{equation}
G_{ab}^{\alpha \beta }\equiv G_{n}^{\alpha \beta }\lambda _{ab}^{n}/2,\ \
G^{2}=G_{\alpha \beta }^{n}G_{n}^{\alpha \beta },\ \ G^{3}=f^{nmd}G_{\alpha
\beta }^{n}G^{n\beta \delta }G_{\delta }^{d\alpha },
\end{equation}%
where $f^{nmd}$ are structure constants of the color group $SU_{c}(3)$.

The $\Pi _{\mu \nu }^{\mathrm{OPE}}(p)$ depends on two important factors:
One of them is the trace term with the heavy quark propagators. The
propagator $S_{Q}^{ab}(x)$ contains the perturbative component and
nonperturbative terms proportional to $g_{s}^{2}G^{2}$ and $g_{s}^{3}G^{3}$
that after sandwiched between vacuum states give rise to well known gluon
condensates. But there also is a term $\sim g_{s}G_{ab}^{\alpha \beta }$ in
the propagator which having multiplied with a similar component of the
second propagator generates additional two-gluon condensate: All such terms
are taken into account.

Another question to be clarified here is connected with the matrix element $%
\langle 0|g_{s}^{2}G^{n\alpha \beta }(x)G^{m\alpha ^{\prime }\beta ^{\prime
}}(0)|0\rangle $. Our treatment of this matrix element is twofold. First, we
replace it by the vacuum condensate $\langle g_{s}^{2}G^{2}\rangle $ keeping
the first term in the Taylor expansion at $x=0$
\begin{equation}
\langle 0|g_{s}^{2}G^{n\alpha \beta }(x)G^{m\alpha ^{\prime }\beta ^{\prime
}}(0)|0\rangle =\frac{\langle g_{s}^{2}G^{2}\rangle }{96}\delta ^{nm}\left[
g^{\alpha \alpha ^{\prime } }g^{\beta \beta ^{\prime }}-g^{\alpha \beta
^{\prime }}g^{\alpha ^{\prime }\beta }\right] .  \label{eq:GluonME1}
\end{equation}%
Terms generated by this way correspond to diagrams in which the gluon
interacts with the QCD vacuum. Alternatively, instead of $\ \langle
0|G^{n\alpha \beta }(x)G^{m\alpha ^{\prime }\beta ^{\prime }}(0)|0\rangle $
we use the full gluon propagator in the $x$-space
\begin{eqnarray}
&&\langle 0|G^{n\alpha \beta }(x)G^{m\alpha ^{\prime }\beta ^{\prime
}}(0)|0\rangle =\frac{\delta ^{nm}}{2\pi ^{2}x^{4}}\left[ g^{\beta \beta
^{\prime }}\left( g^{\alpha \alpha ^{\prime }}-\frac{4x_{\alpha }x_{\alpha
^{\prime }}}{x^{2}}\right) \right.  \notag \\
&&\left. +(\beta ,\beta ^{\prime })\leftrightarrow (\alpha ,\alpha ^{\prime
})-\beta \leftrightarrow \alpha -\beta ^{\prime }\leftrightarrow \alpha
^{\prime }\right] .  \label{eq:GluonME2}
\end{eqnarray}%
Contributions obtained by this manner describe diagrams with full valence
gluon propagator.

The correlator $\Pi _{\mu \nu }^{\mathrm{OPE}}(p)$ is also a sum of terms $%
\sim g_{\mu \nu }$ and $\sim p_{\mu }p_{\nu }$. The amplitude $\Pi _{\mathrm{%
V}}^{\mathrm{OPE}}(p^{2})$ that corresponds to the structure $g_{\mu \nu }$
is employed to find the parameters $m_{\mathrm{V}}$ and $f_{\mathrm{V}}$.
The amplitude $\Pi _{\mathrm{S}}^{\mathrm{OPE}}(p^{2})$ extracted from the
correlation function $\widetilde{\Pi }^{\mathrm{OPE}}(p)=p^{\mu }p^{\nu }\Pi
_{\mu \nu }^{\mathrm{OPE}}(p)/p^{2}$ is convenient to determine SRs for $m_{%
\mathrm{S}} $ and $f_{\mathrm{S}}$.

Having equated the amplitudes $\Pi _{\mathrm{S}}^{\mathrm{OPE}}(p^{2})$ and $%
\Pi _{\mathrm{S}}^{\mathrm{Phys}}(p^{2})$ and carried out the Borel
transformation and continuum subtraction, we find the following sum rules
\begin{equation}
m_{\mathrm{S}}^{2}=\frac{\Pi _{\mathrm{S}}^{\prime }(M^{2},s_{0})}{\Pi _{%
\mathrm{S}}(M^{2},s_{0})}  \label{eq:Mass}
\end{equation}%
and
\begin{equation}
f_{\mathrm{S}}^{2}=\frac{e^{m_{\mathrm{S}}^{2}/M^{2}}}{m_{\mathrm{S}}^{2}}%
\Pi _{\mathrm{S}}(M^{2},s_{0}),  \label{eq:Coupl}
\end{equation}%
where $\Pi _{\mathrm{S}}(M^{2},s_{0})$ is the amplitude $\Pi _{\mathrm{S}}^{%
\mathrm{OPE}}(p^{2})$ obtained after the Borel transformation and continuum
subtraction procedures. Here, $M^{2}$ and $s_{0}$ are the Borel and
continuum subtraction parameters, respectively. In Eq.\ (\ref{eq:Mass}), we
have also introduced the short notation $\Pi _{\mathrm{S}}^{\prime
}(M^{2},s_{0})=d/d(-1/M^{2})\Pi _{\mathrm{S}}(M^{2},s_{0}).$

The amplitude $\Pi _{\mathrm{S}}(M^{2},s_{0})$ has the form
\begin{equation}
\Pi _{\mathrm{S}}(M^{2},s_{0})=\int_{4m_{Q}^{2}}^{s_0}ds\rho _{\mathrm{S}}^{%
\mathrm{OPE}}(s)e^{-s/M^2}+\Pi _{\mathrm{S}}(M^{2}),  \label{eq:CF3}
\end{equation}%
where $\rho _{\mathrm{S}}^{\mathrm{OPE}}(s)$ is the two-point spectral
density. The term $\Pi _{\mathrm{S}}(M^{2})$ stands for nonperturbative
contributions calculated directly from $\Pi _{\mathrm{S}}^{\mathrm{OPE}%
}(p^{2})$. Explicit expressions of the functions $\rho _{\mathrm{S}}^{%
\mathrm{OPE}}(s)$ and $\Pi _{\mathrm{S}}(M^{2})$ are rather cumbersome,
therefore we do not write down them here.

In the present paper $\Pi _{\mathrm{S}}(M^{2},s_{0})$ is computed by taking
into account terms up to dimension $12$. The propagator Eq.\ (\ref{eq:Prop})
contains gluon condensates of different dimensions. The dimension $4$ and $6$
terms proportional to condensates $\langle g_{s}^{2}G^{2}\rangle $ and \ $%
\langle g_{s}^{3}G^{3}\rangle $ appear in final expressions due to existence
of relevant components in the heavy quark propagator. The terms of $8$, $10$
and $12$ dimensions are calculated by means of the factorization hypothesis
of the higher dimensional condensates. But this assumption is not precise
and violates in the case of higher dimensional condensates \cite%
{Ioffe:2005ym}. Nevertheless, in what follows, we neglect uncertainties
generated by this violation because higher dimensional contributions
themselves are small.

The sum rules for $m_{\mathrm{S}}$ and $f_{\mathrm{S}}$ contain, as input
parameters, gluon condensates and the mass of $c$ quark. Below, we list their
numerical values
\begin{eqnarray}
&&\langle \alpha _{s}G^{2}/\pi \rangle =(0.012\pm 0.004)~\mathrm{GeV}^{4},\
\langle g_{s}^{3}G^{3}\rangle =(0.57\pm 0.29)~\mathrm{GeV}^{6}  \notag \\
&&m_{c}(\mu =m_{c})=(1.27\pm 0.02)~\mathrm{GeV},\ m_{b}(\mu
=m_{b})=4.18_{-0.02}^{+0.03}~\mathrm{GeV}.  \label{eq:Param}
\end{eqnarray}%
Above we also present $b$ quark's mass necessary for analysis of the hybrids
$\overline{b}gb$ and $\overline{b}gc$. The $m_{c}$ and $m_{b}$ correspond to
the running masses in the $\overline{\mathrm{MS}}$ scheme at the scales $\mu
=m_{c}$ and $\mu =m_{b}$ \cite{PDG:2022}, respectively. The condensates $%
\langle \alpha _{s}G^{2}/\pi \rangle $ and $\langle g_{s}^{3}G^{3}\rangle $
were extracted from analysis of hadronic processes \cite%
{Shifman:1978bx,Shifman:1978by,Narison:2015nxh}.

Equations\ (\ref{eq:Mass}) and (\ref{eq:Coupl}) depend on the parameters $%
M^{2}$ and $s_{0}$ which should be chosen to meet requirements of SR
analysis. In other words, they have to ensure the dominance of the pole
contribution ($\mathrm{PC}$) in extracted physical quantities. Convergence
of the $\mathrm{OPE}$ and stability of obtained results on $M^{2}$ are among
important constraints as well. To keep under control these features of the
SR computations, we employ
\begin{equation}
\mathrm{PC}=\frac{\Pi (M^{2},s_{0})}{\Pi (M^{2},\infty )},  \label{eq:PC}
\end{equation}%
and%
\begin{equation}
R(M^{2})=\frac{\Pi ^{\mathrm{DimN}}(M^{2},s_{0})}{\Pi (M^{2},s_{0})},
\label{eq:Conv}
\end{equation}%
where $\Pi ^{\mathrm{DimN}}(M^{2},s_{0})=\sum_{\mathrm{N}=8,10,12}\Pi ^{%
\mathrm{DimN}}$ is a sum of last three terms in $\mathrm{OPE}$ which are
proportional to $\langle g_{s}^{2}G^{2}\rangle ^{2}$,\ $\langle
g_{s}^{2}G^{2}\rangle \langle g_{s}^{3}G^{3}\rangle $ and $\langle
g_{s}^{3}G^{3}\rangle ^{2}$, respectively.

\begin{figure}[h!]
\begin{center}
\includegraphics[totalheight=6cm,width=8cm]{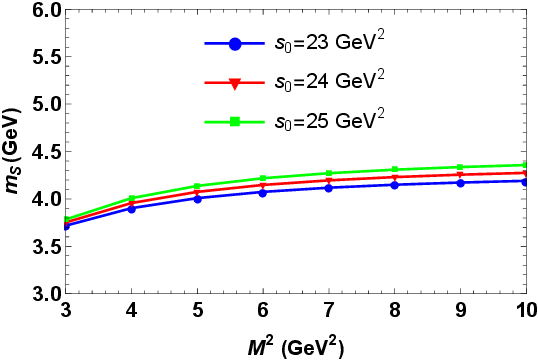} %
\includegraphics[totalheight=6cm,width=8cm]{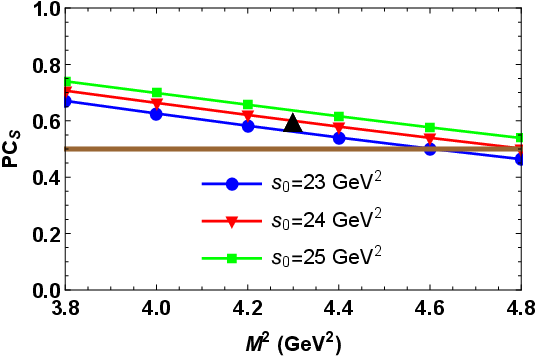}
\end{center}
\caption{Mass $m_{\mathrm{S}}$ of the scalar $J^{\mathrm{PC}}=0^{+-}$
charmonium hybrid as a function of the Borel parameter (left). The pole
contribution $\mathrm{PC_{\mathrm{S}}}$ vs Borel parameter $M^{2}$ at fixed $%
s_{0}$ for the same particle (right). The horizontal line show a region $%
\mathrm{PC}=0.5$. The triangle denote the point, where the mass $m_{\mathrm{S%
}}$ is extracted. }
\label{fig:PC1}
\end{figure}

Now, we concentrate on charmonium hybrids $H_{\mathrm{S}}$ and $H_{\mathrm{V}%
}$. In Fig.\ \ref{fig:PC1} we plot the mass of the hybrid $H_{\mathrm{S}}$
as a function of the Borel parameter $M^{2}=3-10~\mathrm{GeV}^{2}$ at fixed $%
s_{0}$. Our numerical computations prove that in the case of $H_{\mathrm{S}}$
the regions
\begin{equation}
M^{2}\in \lbrack 3.8,4.8]~\mathrm{GeV}^{2},\ s_{0}\in \lbrack 23,25]~\mathrm{%
GeV}^{2},  \label{eq:Wind1}
\end{equation}%
meet constraints of SR analysis. Thus, at $M^{2}=4.8~\mathrm{GeV}^{2}$ and $%
M^{2}=3.8~\mathrm{GeV}^{2}$ on the average in $s_{0}$ the pole contribution
is $\mathrm{PC}_{\mathrm{S}}\approx 0.5$ and $\mathrm{PC}_{\mathrm{S}}$ $%
\approx 0.71$, respectively. At $M^{2}=3.8~\mathrm{GeV}^{2}$ the
nonperturbative contribution is positive and constitutes $<1\%$ of $\ \Pi _{%
\mathrm{S}}(M^{2},s_{0})$. The narrowness of the window for the Borel
parameter is connected with the strong restriction $\mathrm{PC}\geq 0.5$
imposed on the pole contribution. Dependence of $\mathrm{PC}_{\mathrm{S}}$
on the Borel parameter $M^{2}$ is plotted in Fig.\ \ref{fig:PC1}.

The mass $m_{\mathrm{S}}$ and current coupling $f_{\mathrm{S}}$ are
evaluated as mean values of these quantities over the regions Eq.\ (\ref%
{eq:Wind1}): They are equal to
\begin{eqnarray}
m_{\mathrm{S}} &=&(4.06\pm 0.12\pm 0.05\pm 0.01)~\mathrm{GeV},\ \   \notag \\
f_{\mathrm{S}} &=&(2.80\pm 0.30\pm 0.18\pm 0.09)\times 10^{-2}~\mathrm{GeV}%
^{3},  \label{eq:Result1}
\end{eqnarray}%
respectively. The results in Eq.\ (\ref{eq:Result1}) correspond to the SR
predictions at the point $M^{2}=4.3~\mathrm{GeV}^{2}$ and $s_{0}=24~\mathrm{%
GeV}^{2}$. At these values of $M^{2}$ and $s_{0}$ the pole contribution is $%
\mathrm{PC}_{\mathrm{S}}\approx 0.60$, which ensures its dominance in the
obtained results, and proves ground-state character of $H_{\mathrm{S}}$ in a
relevant class of hybrid quarkonia.

Errors of SR computations in Eq.\ (\ref{eq:Result1}) are generated by the
choices of the parameters $M^{2}$ and $s_{0}$, ambiguities in the gluon
condensate $\langle \alpha _{s}G^{2}/\pi \rangle $ and in the mass $m_{c}$,
respectively. The main sources of theoretical errors in present analysis are
the parameters $M^{2}$, $s_{0}$ and the gluon condensate $\langle \alpha
_{s}G^{2}/\pi \rangle $. Uncertainties of the condensate $\langle
g_{s}^{3}G^{3}\rangle $ lead to corrections, which in the cases of $m_{%
\mathrm{S}}$ and $\widetilde{m}_{\mathrm{S}}$ (see, below), for instance,
are equal to $\pm 4\times 10^{-6}$ $\mathrm{GeV}$ and $<|10^{-6}|$ $\mathrm{%
GeV}$, respectively, and therefore can be safely ignored: Throughout this
work, in calculations we use for $\langle g_{s}^{3}G^{3}\rangle $ its
central value. Another sources of possible errors are ones due to scale
dependence of the gluon condensates and $c$-quark mass. But $\langle \alpha
_{s}G^{2}/\pi \rangle $ is the $\mu $-scale independent quantity, and
effects of $m_{c}(\mu )$ and $\langle g_{s}^{3}G^{3}\rangle $ rescaling,
which can be included into uncertainties of these quantities, are small and
can be neglected as well. The mass $m_{\mathrm{S}}$ is shown in Fig.\ \ref%
{fig:Mass1} as functions of the Borel and continuum subtraction parameters.

\begin{figure}[h!]
\begin{center}
\includegraphics[totalheight=6cm,width=8cm]{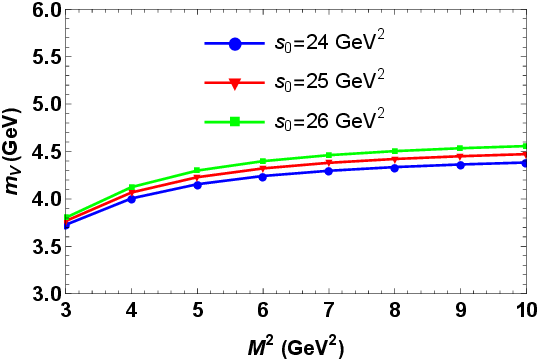} %
\includegraphics[totalheight=6cm,width=8cm]{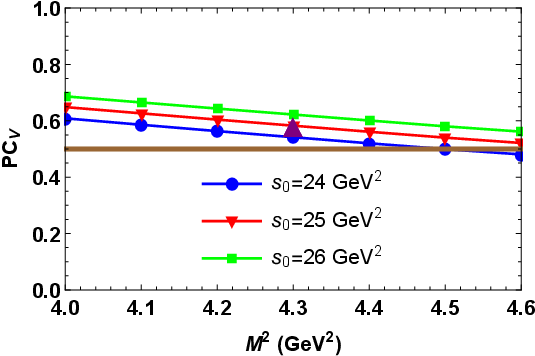}
\end{center}
\caption{The same as in Fig.\ \protect\ref{fig:PC1}, but for the vector
charmonium hybrid $J^{\mathrm{PC}}=1^{--}$. }
\label{fig:PC2}
\end{figure}

\begin{figure}[h!]
\begin{center}
\includegraphics[totalheight=6cm,width=8cm]{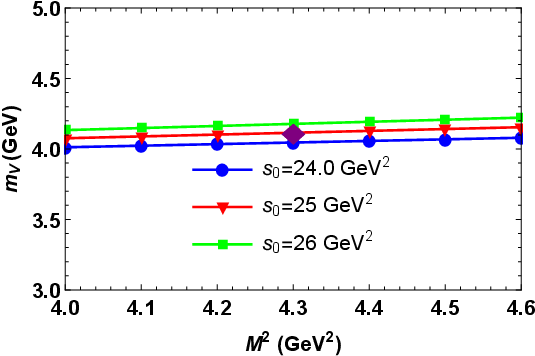} %
\includegraphics[totalheight=6cm,width=8cm]{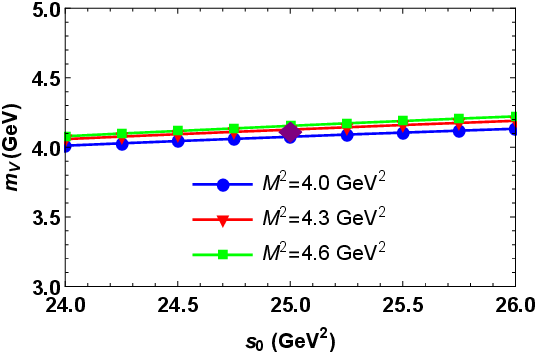}
\end{center}
\caption{The mass $m_{\mathrm{V}}$ of the vector hybrid $H_{\mathrm{V}}$ as
a function of the Borel $M^{2}$ (left panel), and continuum threshold $s_0$
parameters (right panel). The diamonds show positions, where the mass $m_{%
\mathrm{V}}$ has been evaluated.}
\label{fig:Mass2}
\end{figure}

The spectroscopic parameters of the vector hybrid $J^{\mathrm{PC}}=1^{--}$
can be extracted from the sum rules Eqs.\ (\ref{eq:Mass}) and \ (\ref%
{eq:Coupl}) after evident substitutions $\Pi _{\mathrm{S}}(M^{2},s_{0})%
\rightarrow \Pi _{\mathrm{V}}(M^{2},s_{0})$ and $(m_{\mathrm{S}},\ f_{%
\mathrm{S}})\rightarrow (m_{\mathrm{V}},\ f_{\mathrm{V}})$. The graphics for
the mass $m_{\mathrm{V}}$ and $\mathrm{PC}_{\mathrm{V}}$ are depicted in
Fig.\ \ref{fig:PC2}. The mass and current coupling $m_{\mathrm{V}}$ and $f_{%
\mathrm{V}}$ read
\begin{eqnarray}
m_{\mathrm{V}} &=&(4.12\pm 0.11\pm 0.06\pm 0.004)~\mathrm{GeV},\   \notag \\
\ f_{\mathrm{V}} &=&(4.0\pm 0.4\pm 0.2\pm 0.02)\times 10^{-2}~\mathrm{GeV}%
^{3},  \label{eq:Result2}
\end{eqnarray}%
and have been extracted using the parameters%
\begin{equation}
M^{2}\in \lbrack 4,4.6]~\mathrm{GeV}^{2},\ s_{0}\in \lbrack 24,26]~\mathrm{%
GeV}^{2}.
\end{equation}%
The quantities $m_{\mathrm{V}}$ and $f_{\mathrm{V}}$ are effectively
evaluated at $M^{2}=4.3~\mathrm{GeV}^{2}$ and $s_{0}=25~\mathrm{GeV}^{2}$,
where $\mathrm{PC}_{\mathrm{V}}\approx 0.58$. In Fig.\ \ref{fig:Mass2} one
can see dependence of the mass $m_{\mathrm{V}}$ on $M^{2}$ and $s_{0}$.

\begin{figure}[h]
\begin{center}
\includegraphics[totalheight=6cm,width=8cm]{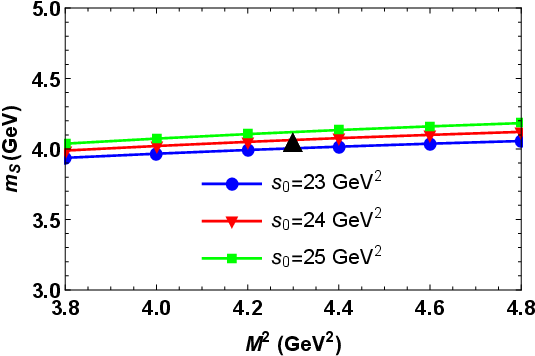} %
\includegraphics[totalheight=6cm,width=8cm]{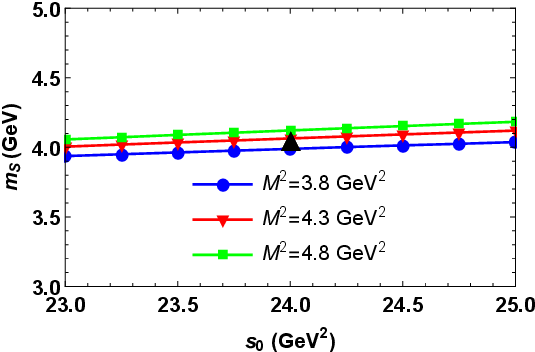}
\end{center}
\caption{Dependence of the mass $m_{\mathrm{S}}$ of the hybrid $H_{\mathrm{S}%
}$ on the Borel $M^{2}$ (left panel), and continuum threshold $s_{0}$
parameters (right panel). The triangles on the plots fix the position of $H_{%
\mathrm{S}}$.}
\label{fig:Mass1}
\end{figure}

Parameters of the charmonium hybrids with different spin-parities $J^{%
\mathrm{PC}}$ are moved to Table \ref{tab:HCharmonia}. The particles are
placed in accordance with the light and heavy hybrid supermultiplet
structures revealed in the MIT bag model \cite%
{Barnes:1982tx,Chanowitz:1982qj} and confirmed by the QCD lattice
simulations \cite{HadronSpectrum:2012gic}. The hybrids $J^{\mathrm{PC}%
}=\left\{ (0,1,2)^{-+};1^{--}\right\} $ composed of $S$-wave color-octet
diquark $c\overline{c}$ and an excited gluon $J^{\mathrm{PC}}=1^{+-}$ form
the light supermultiplet. In this paper, we have restricted ourselves by
analysis of particles $J=0,1$, therefore this multiplet contains only three
states $(\left\{ 0,1)^{-+};1^{--}\right\} $. The heavy multiplet of the
charmonium hybrids built of the $P$-wave diquark $c\overline{c}$ and a gluon
$J^{\mathrm{PC}}=1^{+-}$contains members $J^{\mathrm{PC}}=\left\{
(0,1^{3},2^{2},3)^{+-};(0,1,2)^{++}\right\} $ which in our case reduce to
four states $J^{\mathrm{PC}}=\left\{ (0,1)^{+-};(0,1)^{++}\right\} $.

The mass of the charmonium hybrids with spins $0$ and $1$ change within the
limits $3.56-4.64~\mathrm{GeV}$. The pseudoscalar state $0^{-+}$ has the
mass $(3.56\pm 0.10)~\mathrm{GeV}$ and is a lightest charmonium hybrid. This
result, as well as prediction for the vector particle $1^{-+}$ within errors
agree with ones made in Ref.\ \cite{Chen:2013zia}. The situation around the
state $1^{--}$ is controversial and has been explained above. Our prediction
$m_{\mathrm{V}}=(4.12\pm 0.13)~\mathrm{GeV}$ is compatible with $4.12-4.79~%
\mathrm{GeV}$ of Ref.\ \cite{Qiao:2010zh}, but is considerably larger than $%
(3.36\pm 0.15)~\mathrm{GeV}$ found in Ref.\ \cite{Chen:2013zia}.

In the heavy supermultiplet consisting of four hybrids the masses change
from $(4.06\pm 0.13)~\mathrm{GeV}$ for the scalar $0^{+-}$ to $(4.53\pm
0.17)~\mathrm{GeV}$ for the hybrid $0^{++}$. The masses of the particles $%
0^{++}$ and $1^{++}$ are considerably lower than ones reported in Ref.\ \cite%
{Chen:2013zia}. The heaviest state among considered structures is the
charmonium hybrid bearing the exotic quantum numbers $0^{--}$: Our result
for the mass of this state $(4.64\pm 0.15)~\mathrm{GeV}$ is $0.87~\mathrm{GeV%
}$ smaller than prediction of \cite{Chen:2013zia}.

Table \ref{tab:HCharmonia} also contains information on ordinary $\overline{c%
}c$ mesons to compare a mass hierarchy of normal and hybrid charmonia. As is
seen, states $0^{-+}$ and $1^{--}$ are $\delta _{c}(0^{-+})\approx 0.58~%
\mathrm{GeV}$ and $\delta _{c}(1^{--})\approx 1~\mathrm{GeV}$ above the
normal charmonia $\eta _{c}$ and $J/\psi $. The hybrids $0^{++}$, $1^{+-}$
and $1^{++}$ are heavier than the mesons $\chi _{c0}(1P)$, $h_{c}(1P)$, and $%
\chi _{c1}(1P)$: Their masses overshoot corresponding values approximately
by $\delta _{c}(0^{++})\approx 1.12$, $\delta _{c}(1^{+-})\approx 0.69$, and
$\delta _{c}(1^{++})\approx 0.64~\mathrm{GeV}$, respectively.

For comparison, the charmonium hybrids' parameters obtained in the context
of different method are presented in Table \ref{tab:Comparison}. Besides the
sum rule predictions, we write down there results of BOEFT and QCD lattice
approaches. A relatively nice agreement is observed between the present
analysis and BOEFT from Ref.\ \cite{Soto:2023lbh}. The lattice simulations
generate considerably higher outputs \cite{Cheung:2016bym}. Thus, the mass
splitting between hybrids from the light multiplet and normal charmonia was
found in Ref.\ \cite{Cheung:2016bym} of around $1.2-1.4~\mathrm{GeV}$,
whereas in the heavy supermultiplet this difference amounts to $1.1-1.2~%
\mathrm{GeV}$, which for many positions are higher than our estimates.

\begin{table}[tbp]
\begin{tabular}{|c|c|c|c|c|c|c|}
\hline\hline
$J^{\mathrm{PC}}$ & $M^2~(\mathrm{GeV}^2)$ & $s_0~(\mathrm{GeV}^2)$ & $%
\mathrm{PC} (\%)$ & Mass $(\mathrm{GeV})$ & $f\times 10^{2} (\mathrm{GeV}^3)$
& Mass of $\overline{c}c$ meson $(\mathrm{MeV})$ \\ \hline
$0^{-+}$ & $3.6-4.7$ & $19-20$ & $72-50$ & $3.56(09)(04)(01)$ & $%
3.8(3)(4)(07)$ & $\eta_c: 2983.9(4)$ \\
$1^{-+}$ & $4-5$ & $23-24$ & $69-50$ & $3.93(10)(07)(007)$ & $%
4.60(20)(17)(11)$ & $-$ \\
$1^{--}$ & $4-4.6$ & $24-26$ & $63-50$ & $4.12(11)(06)(004)$ & $%
4.0(4)(2)(02) $ & $J/\psi: 3096.900(6)$ \\ \hline
$0^{+-}$ & $3.8-4.8$ & $23-25$ & $71-50$ & $4.06(12)(05)(01)$ & $%
2.80(30)(18)(09) $ & $-$ \\
$1^{+-}$ & $4-5$ & $25-27$ & $69-50$ & $4.21(15)(07)(004)$ & $%
4.50(60)(04)(14) $ & $h_c(1P): 3525.37(14)$ \\
$0^{++}$ & $4-5$ & $26-28$ & $75-50$ & $4.53(16)(07)(007)$ & $%
2.50(90)(37)(05)$ & $\chi_{c0}(1P): 3414.71(30)$ \\
$1^{++}$ & $4.5-5.1$ & $26-28$ & $70-59$ & $4.15(09)(07)(007)$ & $%
5.80(50)(10)(14)$ & $\chi_{c1}(1P): 3510.67(05)$ \\ \hline
$0^{--}$ & $5-6.5$ & $30-32$ & $70-50$ & $4.64(14)(06)(01)$ & $%
4.90(50)(14)(13)$ & $- $ \\ \hline\hline
\end{tabular}%
\caption{Mass and current coupling of the hybrid charmonia $\overline{c}gc$,
their quantum numbers and input parameters used in calculations. We provide
also the masses of the conventional $\overline{c}c$ mesons \protect\cite%
{PDG:2022}. Errors of calculations and experiments are shown in a compact
form, for example, $3.56(09)(04)(01)$ implies $3.56 \pm 0.09 \pm 0.04 \pm
0.01$.}
\label{tab:HCharmonia}
\end{table}

Information obtained for the masses of the charmonium hybrids allows us to
fix their kinematically allowed two-body strong decay channels. Here, we
consider this problem only qualitatively, without calculation partial widths
of fixed modes. The hybrid structures can decay through open- and
hidden-charm exclusive channels. Decay pattern of the hybrid mesons, their
suppressed or preferable modes, model-dependent selection rules were
investigated in many publications \cite%
{Kou:2005gt,Isgur:1985vy,Close:1994hc,Page:1996rj,Page:1998gz,TarrusCastella:2021pld,Bruschini:2023tmm}%
.

\begin{table}[tbp]
\begin{tabular}{|c|c|c|c|c|}
\hline\hline
$J^{\mathrm{PC}}$ & This work & SR \cite{Chen:2013zia} & BOEFT \cite%
{Soto:2023lbh} & Lattice \cite{Cheung:2016bym} \\ \hline
$0^{-+}$ & $3.56(10)$ & $3.61(21)$ & $3.911(54)$ & $4.279(18)$ \\
$1^{-+}$ & $3.93(12)$ & $3.70(21)$ & $3.963(38)$ & $4.310(23)$ \\
$1^{--}$ & $4.12(13)$ & $3.36(15)$ & $4.011(30)$ & $4.411(17)$ \\ \hline
$0^{+-}$ & $4.06(13)$ & $4.09(23)$ & $4.087(61)$ & $4.437(27)$ \\
$1^{+-}$ & $4.21(17)$ & $4.53(23)$ & $4.235(35)$ & $4.665(53)$ \\
$0^{++}$ & $4.53(17)$ & $5.34(45)$ & $4.486(30)$ & $4.591(46)$ \\
$1^{++}$ & $4.15(11)$ & $5.06(44)$ & $4.145(30)$ & $4.518(35)$ \\ \hline
$0^{--}$ & $4.64(15)$ & $5.51(50)$ & $-$ & $-$ \\ \hline\hline
\end{tabular}%
\caption{Predictions for the masses (in $\mathrm{GeV}$ units) of the hybrid
charmonia $\overline{c}gc$ obtained in different articles. }
\label{tab:Comparison}
\end{table}

The all isoscalar hybrid charmonia considered here, can decay through
two-body processes to standard mesons. For example, even the lightest state $%
0^{-+}$ in $S$-wave transforms to mesons $\eta _{c}f_{0}(500)$. The mass of
the vector hybrid $1^{--}$ makes the $S$-wave hidden-charm channel
\begin{equation}
H_{\mathrm{V}}\rightarrow J/\psi f_{0}(500),\
\end{equation}%
a possible process for this particle. The open-charm $P$-wave decays%
\begin{equation}
H_{\mathrm{V}}\rightarrow D\overline{D},\ D_{0}\overline{D}_{0},\ D_{s}%
\overline{D}_{s}
\end{equation}%
are among kinematically allowed modes of $H_{\mathrm{V}}$ as well.

The hybrid bottomonia $\overline{b}gb$ can be studied in accordance with the
scheme outlined above. The differences here are connected with the mass of $%
b $-quark, and a necessity to choose new parameters $M^{2}$ and $s_{0}$ in
such a way that to satisfy requirements of SR calculations. Our studies
demonstrate that for the bottomonia counterparts of scalar and vector
charmonium hybrids, i,e., for states $0^{+-}$ and $1^{--}$ the Borel and
continuum threshold parameters have to be fixed in the following limits: for
the scalar particle
\begin{equation}
M^{2}\in \lbrack 12,13.5]~\mathrm{GeV}^{2},\ s_{0}\in \lbrack 124,126]~%
\mathrm{GeV}^{2}  \label{eq:Wind2}
\end{equation}%
and for the vector state
\begin{equation}
M^{2}\in \lbrack 12,14]~\mathrm{GeV}^{2},\ s_{0}\in \lbrack 120,125]~\mathrm{%
GeV}^{2}.  \label{eq:Wind2a}
\end{equation}%
The mass and current coupling of the structures $0^{+-}$ and $1^{--}$ are
equal to
\begin{eqnarray}
\widetilde{m}_{\mathrm{S}} &=&(10.12\pm 0.06\pm 0.06\pm 0.01)~\mathrm{GeV},\
\notag \\
\widetilde{f}_{\mathrm{S}} &=&(4.6\pm 0.3\pm 0.7\pm 0.2)\times 10^{-2}~%
\mathrm{GeV}^{3},  \label{eq:Result3}
\end{eqnarray}%
and
\begin{eqnarray}
\widetilde{m}_{\mathrm{V}} &=&(10.41\pm 0.18\pm 0.08\pm 0.012)~\mathrm{GeV},
\notag \\
\ \widetilde{f}_{\mathrm{V}} &=&(12.0\pm 3.0\pm 0.6\pm 0.4)\times 10^{-2}~%
\mathrm{GeV}^{3},  \label{eq:Result4}
\end{eqnarray}%
respectively.

\begin{table}[tbp]
\begin{tabular}{|c|c|c|c|c|c|c|}
\hline\hline
$J^{\mathrm{PC}}$ & $M^2~(\mathrm{GeV}^2)$ & $s_0~(\mathrm{GeV}^2)$ & $%
\mathrm{PC} (\%)$ & Mass $(\mathrm{GeV})$ & $f\times 10^{2} (\mathrm{GeV}^3)$
& Mass of $\overline{b}b$ meson $(\mathrm{MeV})$ \\ \hline
$0^{-+}$ & $11-13$ & $110-120$ & $67-50$ & $9.68(20)(07)(012)$ & $%
9.2(1.9)(0.28)(0.27) $ & $\eta_b: 9398.7(2.0)$ \\
$1^{-+}$ & $12-13$ & $115-120$ & $60-51$ & $9.85(11)(08)(01)$ & $%
8.9(1.0)(0.13)(0.34) $ & $-$ \\
$1^{--}$ & $12-14$ & $120-125$ & $72-51$ & $10.41(18)(08)(012)$ & $%
12.0(3.0)(0.6)(0.4)$ & $\Upsilon(1S): 9460.40(09)(04)$ \\ \hline
$0^{+-}$ & $12-13.5$ & $124-126$ & $66-50$ & $10.12(06)(06)(01)$ & $%
4.6(3)(7)(2) $ & $-$ \\
$1^{+-}$ & $13.5-14.5$ & $130-132$ & $56-50$ & $10.46(06)(07)(01)$ & $%
11.60(60)(13)(42)$ & $h_b(1P): 9899.3(8)$ \\
$0^{++}$ & $13.5-15$ & $125-130$ & $60-50$ & $10.57(08)(08)(02)$ & $%
19.3(1.6)(0.2)(0.5) $ & $\chi_{b0}(1P): 9859.44(42)(31)$ \\
$1^{++}$ & $12.5-14.5$ & $128-130$ & $65-51$ & $10.55(10)(08)(02)$ & $%
13.4(1.0)(0.07)(0.44)$ & $\chi_{b1}(1P): 9892.78(26)(31)$ \\ \hline
$0^{--}$ & $12-14$ & $130-135$ & $76-50$ & $10.51(11)(06)(02)$ & $%
3.90(30)(09)(18)$ & $-$ \\ \hline\hline
\end{tabular}%
\caption{The same as in Table\ \protect\ref{tab:HCharmonia}, but for the
bottomonia hybrids $\overline{b}gb$. The masses of the mesons $\overline{b}b$
are borrowed from Ref.\ \protect\cite{PDG:2022}. }
\label{tab:HBottomonia}
\end{table}

Other $\overline{b}gb$ states are investigated with similar manner: Results
obtained for the hybrid bottomonia are collected in Table\ \ref%
{tab:HBottomonia} and occupy the mass range of $9.68-10.57~\mathrm{GeV}$.
The particles $0^{-+}$ and $1^{-+}$ from the light supermultiplet have the
masses $(9.68\pm 0.21)~\mathrm{GeV}$ and $(9.85\pm 0.14)~\mathrm{GeV}$,
which coincide with or are very close to ones from Ref.\ \cite{Chen:2013zia}%
. But they are considerably lower than predictions of QCD lattice and BOEFT
analyses. In these approaches masses of the hybrids $0^{-+}$ and $1^{-+}$
were found equal to $(10.926\pm 0.018)~\mathrm{GeV}$ and $(10.935\pm 0.018)~%
\mathrm{GeV}$ \cite{Ryan:2020iog}, and $(10.682\pm 0.005)~\mathrm{GeV}$ and $%
(10.686\pm 0.004)~\mathrm{GeV}$ \cite{Soto:2023lbh}, respectively. In our
analysis, the vector hybrid $1^{--}$ has the mass $(10.41\pm 0.20)~\mathrm{%
GeV}$ which agrees with the result of Ref.\ \cite{Qiao:2010zh}. But it is
significantly larger than $(9.70\pm 0.12)~\mathrm{GeV}$ of Ref.\ \cite%
{Chen:2013zia}, and smaller the lattice and BOEFT predictions.

For bottomonium hybrids from the light supermultiplet the lattice
simulations predict the spin-average mass $\simeq 10.938~\mathrm{GeV}$ \cite%
{Ryan:2020iog}. For particles from the heavy supermultiplet this parameter
equals to $10.872~\mathrm{GeV}$. The Born-Oppenheimer approximation
generates the following average masses \cite{Soto:2023lbh}: The $10.686~%
\mathrm{GeV}$ and $10.822~\mathrm{GeV}$ for the light and heavy
supermultiplets, respectively. One of features of the lattice and BOEFT
spectra is a mass degeneration between the light and heavy supermultiplets.
Comparing these data with our results from Table\ \ref{tab:HBottomonia} and
corresponding spin-averages $9.98~\mathrm{GeV}$ and $10.425~\mathrm{GeV}$ in
the multiplets, we note approximately $1~\mathrm{GeV}$ and $0.4~\mathrm{GeV}$
mass gaps between SR and lattice predictions. In other words, for the hybrid
bottomonia the sum rule calculations give an increasing mass spectrum (see Table \ref{tab:Comparison1}).

\begin{table}[tbp]
\begin{tabular}{|c|c|c|c|c|}
\hline\hline
$J^{\mathrm{PC}}$ & This work & SR \cite{Chen:2013zia} & BOEFT \cite%
{Soto:2023lbh} & Lattice \cite{Ryan:2020iog} \\ \hline
$0^{-+}$ & $9.68(21)$ & $9.68(29)$ & $10.682(5)$ & $10.926(18)$ \\
$1^{-+}$ & $9.85(14)$ & $9.79(22)$ & $10.686(4)$ & $10.935(18)$ \\
$1^{--}$ & $10.41(20)$ & $9.70(12)$ & $10.6902(30)$ & $10.952(24)$ \\ \hline
$0^{+-}$ & $10.12(09)$ & $10.17(22)$ & $10.756(5)$ & $10.935(41)$ \\
$1^{+-}$ & $10.46(09)$ & $10.70(53)$ & $10.759(4)$ & $11.062(35)$ \\
$0^{++}$ & $10.57(12)$ & $11.20(48)$ & $11.012(3)$ & $-$ \\
$1^{++}$ & $10.55(13)$ & $11.09(60)$ & $10.761(3)$ & $10.921(55)$ \\ \hline
$0^{--}$ & $10.51(13)$ & $11.48(75)$ & $-$ & $-$ \\ \hline\hline
\end{tabular}%
\caption{Masses (in $\mathrm{GeV}$ units) of the bottomonium hybrids $%
\overline{b}gb$ extracted in the framework of SR, BOEFT and lattice methods.
}
\label{tab:Comparison1}
\end{table}

The two-body strong decay channels of the $\overline{b}gb$ hybrids is
analyzed based on our estimations. The masses of the hybrids $0^{-+}$ and $%
1^{-+}$ are below thresholds for production of open- and hidden-bottom
two-meson pairs. This conclusion is correct even for upper limits of their
masses. The vector hybrid $1^{--}$ can easily decay in $S$-wave to mesons $%
\Upsilon (1S)f_{0}(500)$ . The scalar particle $0^{+-}$ has the mass $%
(10.12\pm 0.09)~\mathrm{GeV}$ and decays in $P$-wave to a pair $\Upsilon
(1S)f_{0}(500)$. Kinematically allowed two-body decay channels of the
remaining heavy charmonium and bottomonium hybrids can be found in Ref.\
\cite{Chen:2013eha} by taking into account the masses of these particles
found in the present work.


\section{The heavy hybrid mesons $\overline{b}gc$}

\label{sec:Hmesons}
In this section, we consider the heavy hybrid mesons $\overline{b}gc$ and
evaluate masses of these structures with the spin-parities $J^{\mathrm{P}%
}=0^{+},\ 0^{-},\ 1^{+}$, and $1^{-}$. Relevant currents can easily be
obtained from Eqs.\ (\ref{eq:C1}), (\ref{eq:C2}), (\ref{eq:C3}) and (\ref%
{eq:C4}). It is clear that charged hybrids $\overline{b}gc$ can not be
classified by $\mathrm{C}$-parity. But to distinguish particles explored by
means of different currents, we keep formally in parenthesis the $\mathrm{C}$%
-parity of original currents. The current that corresponds to the hybrids $%
J^{\mathrm{P}}=0^{+(-)}$ and $1^{-(-)}$ is%
\begin{equation}
\widetilde{J}_{\mu }^{1}(x)=g_{s}\overline{b}_{a}(x)\gamma ^{\alpha }\gamma
_{5}\frac{{\lambda }_{ab}^{n}}{2}\widetilde{G}_{\mu \alpha }^{n}(x)c_{b}(x).
\label{eq:C5}
\end{equation}%
The scalar and vector particles with $J^{\mathrm{P}}=0^{+(+)}$ and $1^{-(+)}$
can be investigated using the current
\begin{equation}
\widetilde{J}_{\mu }^{2}(x)=g_{s}\overline{b}_{a}(x)\gamma ^{\alpha }\frac{{%
\lambda }_{ab}^{n}}{2}G_{\mu \alpha }^{n}(x)c_{b}(x).  \label{eq:C6}
\end{equation}%
The particles $0^{-(-)}\ $\ and $1^{+(-)}$ are described by the current
\begin{equation}
\widetilde{J}_{\mu }^{3}(x)=g_{s}\overline{b}_{a}(x)\gamma ^{\alpha }\gamma
_{5}\frac{{\lambda }_{ab}^{n}}{2}G_{\mu \alpha }^{n}(x)c_{b}(x).
\label{eq:C7}
\end{equation}%
The current
\begin{equation}
\widetilde{J}_{\mu }^{4}(x)=g_{s}\overline{b}_{a}(x)\gamma ^{\alpha }\frac{{%
\lambda }_{ab}^{n}}{2}\widetilde{G}_{\mu \alpha }^{n}(x)c_{b}(x),
\label{eq:C8}
\end{equation}%
corresponds to the pseudoscalar and axial-vector hybrids $J^{\mathrm{P}%
}=0^{-(+)}\ $\ and $1^{+(+)}$.

The correlation functions $\widetilde{\Pi }_{\mu \nu }(p)$ to be studied in
these cases are given by Eq.\ (\ref{eq:CF1}) after replacement $J_{\mu
}(x)\rightarrow \widetilde{J}_{\mu }(x)$. Treatment of the relevant
correlators does not differ considerably from the analysis presented in the
previous section. Differences appear only in the QCD sides of \ SRs. For
instance, in the case of the current $\widetilde{J}_{\mu }^{1}(x)$ the
correlator $\widetilde{\Pi }_{\mu \nu }^{\mathrm{OPE}}(p)$ is determined by
the formula
\begin{eqnarray}
\widetilde{\Pi }_{\mu \nu }^{\mathrm{OPE}}(p) &=&\frac{i\varepsilon _{\mu
\theta \alpha \beta }\varepsilon _{\nu \delta \alpha ^{\prime }\beta
^{\prime }}}{4}\int d^{4}xe^{ipx}\frac{{\lambda }_{ab}^{n}{\lambda }%
_{a^{\prime }b^{\prime }}^{m}}{4}\mathrm{Tr}\left[ S_{b}^{a^{\prime
}a}(-x)\gamma ^{\theta }\gamma _{5}S_{c}^{bb^{\prime }}(x)\gamma ^{\delta
}\gamma _{5}\right]  \notag \\
&&\times \langle 0|g_{s}^{2}G^{n\alpha \beta }(x)G^{m\alpha ^{\prime }\beta
^{\prime }}(0)|0\rangle .
\end{eqnarray}

Parameters of the bottom-charm hybrids obtained using these interpolating
currents are presented in Table \ref{tab:Hbgc}. As samples, in Fig.\ \ref%
{fig:Mass4} we show masses of the pseudoscalar and vector particles $%
0^{-(+)} $ and $1^{-(+)}$.

\begin{table}[tbp]
\begin{tabular}{|c|c|c|c|c|c|c|}
\hline\hline
$J^{\mathrm{PC}}$ & $M^2~(\mathrm{GeV}^2)$ & $s_0~(\mathrm{GeV}^2)$ & $%
\mathrm{PC} (\%)$ & Mass $(\mathrm{GeV})$ & $f\times 10^{2} (\mathrm{GeV}^3)$
& Mass of $\overline{b}c$ meson $(\mathrm{MeV})$ \\ \hline
$0^{-(+)}$ & $7-8$ & $55-57$ & $69-51$ & $6.55(08)(06)(009)(007)$ & $%
4.9(4)(3)(09)(04)$ & $B_c(1S): 6274.47(27)(17)$ \\
$1^{-(+)}$ & $6.8-7.8$ & $55-60$ & $62-50$ & $6.63(14)(07)(007)(007)$ & $%
4.50(70)(18)(09)(04)$ & $-$ \\
$1^{-(-)}$ & $7-8.3$ & $62-65$ & $71-55$ & $7.01(13)(08)(006)(002)$ & $%
4.80(60)(05)(11)(07)$ & $B_{c}^{\ast}[B_c(1^{3}S_{1})]: 6338$ \\
$0^{+(-)}$ & $8.5-9.4$ & $58-60$ & $61-53$ & $6.93(07)(05)(006)(006)$ & $%
2.80(20)(19)(06)(05) $ & $-$ \\
$1^{+(-)}$ & $7.6-8.6$ & $65-67$ & $61-50$ & $7.17(09)(07)(005)(004)$ & $%
5.70(50)(08)(13)(07)$ & $B_c(1^{1}P_{1}): 6750$ \\
$0^{+(+)}$ & $8-10$ & $65-67$ & $70-50$ & $7.03(12)(08)(008)(005)$ & $%
8.1(8)(2)(1)(07) $ & $B(1^{3}P_{0}): 6706$ \\
$1^{+(+)}$ & $8-9.5$ & $66-67$ & $65-50$ & $7.12(09)(08)(007)(004)$ & $%
7.40(50)(06)(14)(08) $ & $B(1^{3}P_{1}): 6741$ \\
$0^{-(-)}$ & $7.8-8.8$ & $66-67$ & $62-50$ & $7.19(06)(06)(007)(006)$ & $%
4.00(20)(13)(08)(06)$ & $-$ \\ \hline\hline
\end{tabular}%
\caption{The parameters of the bottom-charm hybrid mesons $\overline{b}gc$.
The errors are generated by ambiguities in $(M^2, s_0)$, $\langle \protect%
\alpha _{s}G^{2}/\protect\pi \rangle$, $m_b$, and $m_c$, respectively. The
last column contains masses of the mesons $\overline{b}c$. As the mass of
the meson $B_{c}^{+}(1S)$, we use its experimental value \protect\cite%
{PDG:2022}. The masses of the remaining $\overline{b}c$ mesons are
theoretical predictions obtained by means of the relativized quark model
\protect\cite{Godfrey:2004ya}. }
\label{tab:Hbgc}
\end{table}

\begin{figure}[h]
\begin{center}
\includegraphics[totalheight=6cm,width=8cm]{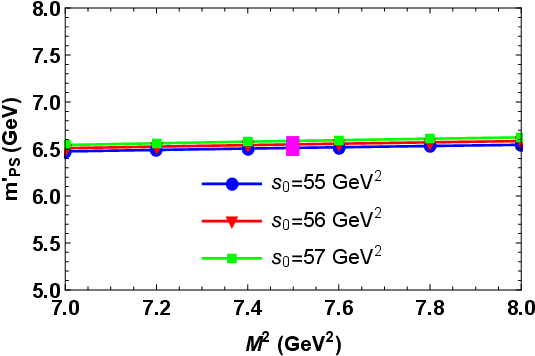} %
\includegraphics[totalheight=6cm,width=8cm]{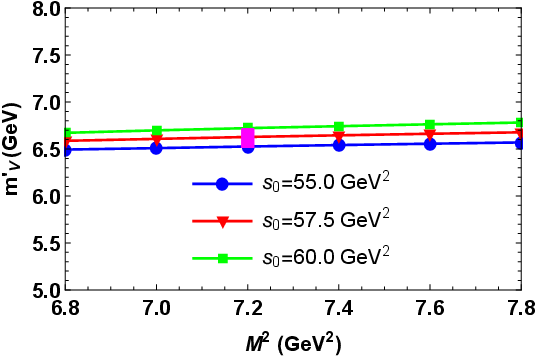}
\end{center}
\caption{Masses of the pseudoscalar $0^{-(+)}$ (left panel) and vector $%
1^{-(+)}$ (right panel) hybrids $\overline{b}gc$ as functions of the Borel
parameter. The rectangles on plots note the hybrid masses.}
\label{fig:Mass4}
\end{figure}

The masses of the hybrid mesons $\overline{b}gc$ change in the interval $%
6.55-7.19~\mathrm{GeV}$. The average masses of hybrids conditionally
belonging to the light multiplet $6.73~\mathrm{GeV}$ is close to $6.9~%
\mathrm{GeV}$ from Ref.\ \cite{Chen:2013eha} and only $0.4~\mathrm{GeV}$
higher than a mass average of two mesons $B_{c}$ and $B_{c}^{\ast }$. But
our predictions for the particles from the heavy multiplet are considerably
smaller than results presented there. The spectrum of the hybrid mesons $%
\overline{b}gc$ evaluated in the present work is very smooth relative to
predictions of Ref.\ \cite{Chen:2013eha}, where the masses vary inside
limits $6.83-8.48~\mathrm{GeV}$.

The possible two-body strong decay modes of the hybrids $\overline{b}gc$ are
determined by thresholds for production of meson pairs with appropriate
quantum numbers. The hybrids $0^{-(+)}\ $\ and $1^{-(+)}$ are light
particles, and seem are stable against strong two-body decays to standard
heavy mesons. The next state in the table $1^{-(-)}$ has allowed channels $%
1^{-(-)}\rightarrow B_{c}^{\ast +}f_{0}(500)$ and $B_{c}^{+}\eta $ which are
its $S$- and $P$-wave decay modes, respectively. The hybrid $0^{+(-)}$
transforms to the meson pairs $B_{c}^{+}\eta $ and $B_{c}^{\ast +}f_{0}(500)$%
. The decay channels of the particles $1^{+(-)}$ and $1^{+(+)}$ are
processes $1^{+(-)}\rightarrow B_{c}^{\ast +}\eta $, $B_{c}^{+}f_{0}(500)$, $%
B_{c}^{+}\omega (782)$ and $1^{+(+)}\rightarrow B_{c}^{+}f_{0}(500)$. The
particles $0^{+(+)}$ and $0^{-(-)}$ can decay through the modes $%
0^{+(+)}\rightarrow B_{c}^{+}\eta $, $B^{+}D^{0}$ and $0^{-(-)}\rightarrow
B_{c}^{+}f_{0}(500)$, $B_{c}^{\ast +}\eta $.


\section{Conclusions}

\label{sec:Dis} 

In this article, we have investigated the scalar, pseudoscalar, vector and
axial-vector charmonium and bottomonium hybrids with positive and negative $%
C $-parities in the context of QCD two-point sum rule approach and extracted
their masses and current couplings. We have also studied the hybrid mesons $%
\overline{b}gc$ with the spin-parities $J^{\mathrm{P}}=0^{+},\ 0^{-},\ 1^{+}$%
, and $1^{-}$.

In our calculations, we have taken into account nonperturbative terms up to
dimension $12$. The higher dimensional terms despite the fact that are
numerically small may be important to improve the stability of SR
calculations. We also imposed the strong restriction on the pole
contribution $\mathrm{PC}\geq 0.5$ which is necessary to extract reliable
predictions for physical quantities under consideration.

Results obtained for the masses of these structures were used to reveal
their possible two-body strong decay channels. It has been demonstrated that
all charmonium hybrids are unstable against strong decays. The $\overline{c}%
gc$ hybrid $1^{--}$ which was predicted to be stable \cite{Chen:2013zia}, in
our case has the mass $m_{\mathrm{V}}=(4.12\pm 0.11)~\mathrm{GeV}$ and
decays through open- and hidden-charm channels to different meson pairs.

In the class of the hybrid bottomonia only the particles $0^{-+}$ and $1^{-+}
$ with the masses $(9.68\pm 0.21)~\mathrm{GeV}$ and $(9.85\pm 0.14)~\mathrm{%
GeV}$ are strong-interaction stable structures, because their masses are
below thresholds for production of open- and hidden-bottom meson pairs. The
exotic mesons $\overline{b}gc$ with quantum numbers $0^{-(+)}\ $\ and $%
1^{-(+)}$ are also stable against strong two-body decays to standard heavy
mesons.

Comparing our results with predictions of Refs.\ \cite%
{Chen:2013zia,Chen:2013eha}, we see that there are differences between them.
The discrepancies are essential for hybrids from the heavy multiplets,
although large uncertainties in extracted masses create overlapping regions
for some of particles. In our view, such deviations are presumably connected
with the requirement $\mathrm{PC}\geq 0.5$ imposed on the pole contributions
in the present analysis.

A nice agreement is achieved with BOEFT results for the charmonium hybrids.
In the case of $\overline{b}gb$ mesons this effective field theory leads to
predictions, especially for the light multiplet, which are higher than ours.
The largest outputs for parameters of the heavy hybrids are generated by the
lattice simulations.

The sum rule analysis performed in the present work and comparisons with
results of alternative methods are important to shed light on properties of
the exotic hybrid mesons. The predictions obtained for the spectroscopic
parameters of the hybrids can be used in investigations of strong and
electroweak decays of these particles as well as to study their interactions
with other hadrons.


\section*{ACKNOWLEDGMENTS}

H.~Sundu is thankful to Scientific and Technological Research Council of T%
\"{u}rkiye (TUBITAK) for the financial support provided under the Grant No.
123F197. K. Azizi is thankful to Iran National Science Foundation (INSF) for
the partial financial support provided under the elites Grant No. 4025036.


\begin{thebibliography}{999}

\bibitem{LHCb:2020bwg} R.~Aaij \textit{et al.} (LHCb Collaboration),
Sci.\ Bull. \textbf{65}, 1983 (2020).


\bibitem{Bouhova-Thacker:2022vnt} E.~Bouhova-Thacker (ATLAS Collaboration),
PoS \textbf{ICHEP2022}, 806 (2022).


\bibitem{CMS:2023owd} A.~Hayrapetyan, \textit{et al.} (CMS Collaboration)
arXiv:2306.07164 [hep-ex].


\bibitem{LHCb:2015yax} R.~Aaij \textit{et al.} (LHCb Collaboration),
Phys.\ Rev. Lett.\ \textbf{115}, 072001 (2015).


\bibitem{LHCb:2019kea} R.~Aaij \textit{et al.} (LHCb Collaboration),
Phys.\ Rev. Lett.\ \textbf{122}, 222001 (2019).


\bibitem{LHCb:2020jpq} R.~Aaij \textit{et al.} (LHCb Collaboration),
Sci.\ Bull. \textbf{66}, 1278 (2021). 


\bibitem{BESIII:2022riz} M.~Ablikim \textit{et al.} (BESIII Collaboration),
Phys.\ Rev. Lett.\ \textbf{129}, 192002 (2022)[E.\textbf{130}, 159901
(2023)]. 


\bibitem{Chen:2022qpd} H.~X.~Chen, N.~Su, and S.~L.~Zhu,
Chin.\ Phys.\ Lett.\ \textbf{39}, 051201 (2022).


\bibitem{Qiu:2022ktc} L.~Qiu, and Q.~Zhao,
Chin.\ Phys.\ Lett.\ \textbf{46}, 051001 (2022).


\bibitem{Shastry:2022mhk} V.~Shastry, C.~S.~Fischer, and F.~Giacosa,
Phys.\ Lett.\ B \textbf{834}, 137478 (2022). 


\bibitem{Kou:2005gt} E.~Kou, and O.~Pene,
Phys.\ Lett.\ B \textbf{631}, 164 (2005). 


\bibitem{Olsen:2017bmm} S.~L.~Olsen, T.~Skwarnicki, and D.~Zieminska,
Rev.\ Mod.\ Phys.\ \textbf{90}, 015003 (2018). 


\bibitem{Brambilla:2022hhi} N.~Brambilla, W.~K.~Lai, A.~Mohapatra, and
A.~Vairo, 
Phys.\ Rev.\ D \textbf{107}, 054034 (2023). 


\bibitem{Engler:1965zz} A.~Engler, H.~E.~Fisk, R.~w.~Kramer, C.~M.~Meltzer,
and J.~B.~Westgard, 
Phys.\ Rev. Lett.\ \textbf{15}, 224 (1965).


\bibitem{Niiyama:2008rt} M.~Niiyama \textit{et al.},
Phys.\ Rev. C \textbf{78}, 035202 (2008). 


\bibitem{HADES:2012csk} G.~Agakishiev \textit{et al.} (HADES Collaboration),
Phys.\ Rev. C \textbf{87}, 025201 (2013). 


\bibitem{CLAS:2013rxx} K.~Moriya \textit{et al.} (CLAS Collaboration),
Phys.\ Rev. C \textbf{88}, 045201 (2013) Addendum:[Phys.\ Rev. C \textbf{88}%
, 049902 (2013)]. 


\bibitem{Jaffe:1975fd} R.~L.~Jaffe and K.~Johnson,
Phys.\ Lett.\ B \textbf{60}, 201 (1976).


\bibitem{Horn:1977rq} D.~Horn, and L.~Mandula,
Phys.\ Rev.\ D \textbf{17}, 898 (1978).


\bibitem{Tanimoto:1982wy} M.~Tanimoto,
Phys.\ Rev.\ D \textbf{27}, 2648 (1983).


\bibitem{Barnes:1982tx} T.~Barnes, F.~E.~Close, and F.~de Viron,
Nucl.\ Phys.\ B \textbf{224}, 241 (1983).


\bibitem{Chanowitz:1982qj} M.~S.~Chanowitz, and S.~R.~Sharpe,
Nucl.\ Phys.\ B \textbf{222}, 211 (1983).


\bibitem{Isgur:1985vy} N.~Isgur, R.~Kokoski, and J.~Paton
Phys.\ Rev.\ Lett.\ \textbf{54}, 869 (1985). 


\bibitem{Deviron:1984svx} F.~de Viron, and J.~Govaerts,
Phys.\ Rev.\ Lett.\ \textbf{53}, 2207 (1984).


\bibitem{Govaerts:1984hc} J.~Govaerts, L.~J.~Reinders, H.~R.~Rubinstein, and
J.~Weyers, 
Nucl.\ Phys.\ B \textbf{258}, 215 (1985).


\bibitem{Govaerts:1985fx} J.~Govaerts, L.~J.~Reinders, and J.~Weyers,
Nucl.\ Phys.\ B \textbf{262}, 575 (1985).


\bibitem{Close:1994hc} F.~E.~Close, and P.~R.~Page,
Nucl.\ Phys.\ B \textbf{443}, 233 (1995). 


\bibitem{Close:1994pr} F.~E.~Close, and P.~R.~Page,
Phys.\ Rev.\ D \textbf{52}, 1706 (1995). 


\bibitem{Page:1996rj} P.~R.~Page,
Phys.\ Lett.\ B \textbf{402}, 183 (1997). 


\bibitem{Page:1998gz} P.~R.~Page, E.~S.~Swanson, and A.~P.~Szczepaniak,
Phys.\ Rev.\ D \textbf{59}, 034016 (1999). 


\bibitem{Zhu:1998sv} S.~L.~Zhu,
Phys.\ Rev.\ D \textbf{60}, 014008 (1999). 


\bibitem{Narison:2009vj} S.~Narison, 
Phys.\ Lett.\ B \textbf{675},319 (2009). 


\bibitem{Qiao:2010zh} C.~F.~Diao, L.~Tang, G.~Hao, and X.~Q.~Li,
J.~Phys.~G \textbf{39}, 015005 (2012). 


\bibitem{Harnett:2012gs} D.~Harnett, R.~T.~Kleiv, T.~G.~Steele, and
H.~y.~Jin,
J.~Phys.~G \textbf{39}, 125003 (2012). 


\bibitem{HadronSpectrum:2012gic} L.~Liu \textit{et al.} (Hadron Spectrum
Collaboration),
JHEP \textbf{07}, 126 (2012). 


\bibitem{Chen:2013zia} W.~Chen, R.~T.~Kleiv, T.~G.~Steele, B.~Bulthuis,
D.~Harnett, T.~Ho, T.~Richards, and S.~L.~Zhu,
JHEP \textbf{09}, 019 (2013). 


\bibitem{Chen:2013eha} W.~Chen, T.~G.~Steele, and S.~L.~Zhu,
J.\ Phys.\ G \textbf{41}, 025003 (2014). 


\bibitem{Cheung:2016bym} G.~K.~C.~Cheung \textit{et al.} (Hadron Spectrum
Collaboration), 
JHEP \textbf{12}, 089 (2016). 


\bibitem{Azizi:2017xyx} K.~Azizi, B.~Barsbay, and H.~Sundu,
Eur.\ Phys.\ J.\ Plus\ \textbf{133}, 121 (2018).


\bibitem{Huang:2014hya} Z.~R.~Huang, H.~Y.~Jin, and Z.~F.~Zhang
JHEP \textbf{04}, 004 (2015). 


\bibitem{Palameta:2018yce} A.~Palameta, D.~Harnett, and T.~G.~Steele,
Phys.\ Rev.\ D \textbf{98}, 074014 (2018). 


\bibitem{Miyamoto:2019oin} T.~Miyamoto and S.~Yasui,
Phys.\ Rev.\ D \textbf{99}, 094015 (2019). 


\bibitem{Brambilla:2018pyn} N.~Brambilla, W.~K.~Lai, J.~Segovia, J.~Tarrus
Castella, and A.~Vairo, 
Phys.\ Rev.\ D \textbf{99}, 014017 (2019)[Erratum: Phys.\ Rev.\ D \textbf{101%
}, 099902 (2020)]. 


\bibitem{Ryan:2020iog} S.~M.~Ryan, and D.~J.~Wilson,
JHEP \textbf{02}, 214 (2021). 


\bibitem{TarrusCastella:2021pld} J.~Tarr\'us Castell\`a and E.~Passemar,
Phys.\ Rev.\ D \textbf{104}, 034019 (2021). 


\bibitem{Woss:2020ayi} A.~J.~Woss \textit{et al.} (Hadron Spectrum
Collaboration), 
Phys.\ Rev.\ D \textbf{103}, 054502 (2021). 


\bibitem{Barsbay:2022gtu} B.~Barsbay, K.~Azizi, and H.~Sundu,
Eur.\ Phys.\ J.\ C \textbf{82}, 1138 (2022). 


\bibitem{Barsbay:2024vjt} B.~Barsbay, K.~Azizi, and H.~Sundu,
arXiv:2402.19006 [hep-ph].


\bibitem{Tang:2021zti} C.~M.~Tang, Y.~C.~Zhao, and L.~Tang,
Phys.\ Rev.\ D \textbf{105}, 114004 (2022). 


\bibitem{Chen:2022isv} F.~Chen, X.~Jiang, Y.~Chen, M.~Gong, Z.~Liu, C.~Shi,
and W.~Sun, 
Phys.\ Rev.\ D \textbf{107}, 054511 (2023). 


\bibitem{Soto:2023lbh} J.~Soto, and S.~T.~Valls,
Phys.\ Rev.\ D \textbf{108}, 014025 (2023). 


\bibitem{Bruschini:2023tmm} R.~Bruschini,
Phys.\ Rev.\ D \textbf{109}, L031501 (2024). 


\bibitem{Shifman:1978bx} M.~A.~Shifman, A.~I.~Vainshtein and V.~I.~Zakharov,
Nucl.\ Phys.\ B \textbf{147}, 385 (1979).


\bibitem{Shifman:1978by} M.~A.~Shifman, A.~I.~Vainshtein and V.~I.~Zakharov,
Nucl.\ Phys.\ B \textbf{147}, 448 (1979).


\bibitem{Nielsen:2009uh} M.~Nielsen, F.~S.~Navarra, and S.~H.~Lee,
Phys.\ Rept.\ \textbf{497}, 41 (2010). 


\bibitem{Albuquerque:2018jkn} R.~M.~Albuquerque, J.~M.~Dias,
K.~P.~Khemchandani, A.~Martinez Torres, F.~S.~Navarra, M.~Nielsen and
C.~M.~Zanetti, 
J.\ Phys.\ G \textbf{46}, 093002 (2019).


\bibitem{Agaev:2020zad} S.~S.~Agaev, K.~Azizi, and H.~Sundu,
Turk.\ J.\ Phys.\ \textbf{44}, 95 (2020). 


\bibitem{Ioffe:2005ym} B.~L.~Ioffe, 
Prog.\ Part.\ Nucl.\ Phys.\ \textbf{56}, 232 (2006).


\bibitem{PDG:2022} R.~L.~Workman \textit{et al.} [Particle Data Group],
Prog.\ Theor.\ Exp.\ Phys.\ \textbf{2022}, 083C01 (2022).


\bibitem{Narison:2015nxh} S.~Narison,
Nucl.\ Part.\ Phys.\ Proc.\ \textbf{270-272}, 143 (2016).


\bibitem{Godfrey:2004ya} S.~Godfrey,
Phys.\ Rev.\ D \textbf{70}, 054017 (2004). 
\end{thebibliography}
\end{document}